\documentclass[aps,pra,twocolumn,eqsecnum,nofootinbib,longbibliography,superscriptaddress]{revtex4-2}

\usepackage{float}
\usepackage{amsmath}
\usepackage{amssymb}
\usepackage{graphicx}
\usepackage{mathrsfs}
\usepackage{color}
\usepackage{lipsum}
\usepackage[procnames]{listings}
\usepackage{layouts}
\usepackage{epstopdf}
\usepackage{dsfont}
\usepackage{bbm}
\counterwithout{equation}{section}
\usepackage[scr=boondox]{mathalfa}
\usepackage{xcolor}
\usepackage{hyperref}

\newcommand{\ket}[1]{{\left\vert{#1}\right\rangle}}

\DeclareUnicodeCharacter{2009}{\,} 

\begin{document}
	
\title{Hierarchies of quantum non-Gaussian coherences\\for bosonic systems: a theoretical study}

	\author{Luk\'a\v s Lachman}
	\email[email: ]{lachmanp@optics.upol.cz}
	\affiliation{Laboratoire Kastler Brossel, Sorbonne Universit\'e, CNRS, ENS-Universit\'e PSL, Coll\`ege de France, 4 Place Jussieu, 75005 Paris, France}
	\affiliation{Department of Optics, Palack\'y University, 17. Listopadu 12, 771 46 Olomouc, Czech Republic}
	\author{Beate E. Asenbeck} 
	\affiliation{Laboratoire Kastler Brossel, Sorbonne Universit\'e, CNRS, ENS-Universit\'e PSL, Coll\`ege de France, 4 Place Jussieu, 75005 Paris, France}
	\author{Ambroise Boyer}
	\affiliation{Laboratoire Kastler Brossel, Sorbonne Universit\'e, CNRS, ENS-Universit\'e PSL, Coll\`ege de France, 4 Place Jussieu, 75005 Paris, France}
	\author{Priyanka Giri}
	\affiliation{Laboratoire Kastler Brossel, Sorbonne Universit\'e, CNRS, ENS-Universit\'e PSL, Coll\`ege de France, 4 Place Jussieu, 75005 Paris, France}
	\author{Alban Urvoy}
	\affiliation{Laboratoire Kastler Brossel, Sorbonne Universit\'e, CNRS, ENS-Universit\'e PSL, Coll\`ege de France, 4 Place Jussieu, 75005 Paris, France}
	\author{Julien Laurat}
	\affiliation{Laboratoire Kastler Brossel, Sorbonne Universit\'e, CNRS, ENS-Universit\'e PSL, Coll\`ege de France, 4 Place Jussieu, 75005 Paris, France}
	\author{Radim Filip}
	\email[email: ]{filip@optics.upol.cz}
	\affiliation{Department of Optics, Palack\'y University, 17. Listopadu 12, 771 46 Olomouc, Czech Republic}

\begin{abstract}
Quantum coherence in bosonic systems is a fundamental resource for quantum technology applications. In this work, we introduce a framework for analyzing coherence in the Fock-state basis, utilizing context-dependent certification to reveal the quantum non-Gaussian nature of the tested coherence. Rather than relying on global coherence measures, our approach targets specific aspects of coherence, enabling a tailored hierarchical classification. We derive and compare two distinct hierarchies, each representing a different context for coherence in bosonic systems. Motivated by current advancements in optical quantum state engineering, we assess the feasibility and depth of these hierarchies under conditions of loss and thermal noise. The methodology introduced here is versatile and can be extended to multi-state and multi-mode coherences, making it adaptable to a wide range of experimental scenarios and platforms.
\end{abstract}	
	
\maketitle
	
\section{Introduction}

Quantum non-Gaussianity represents a pivotal resource in quantum science \cite{Mattia}, yet its inherent diversity raises fundamental questions, particularly in the classification and analysis of quantum states. To address this, quantum non-Gaussianity witnesses based on Fock-state probabilities have been developed  \cite{Filip2011} and applied across various experimental settings \cite{Jezek2012,Straka2014,Straka2018}. Additionally, a hierarchical framework for identifying genuine quantum non-Gaussian features in states approximating ideal Fock states has been introduced \cite{Lachman2019} and experimentally validated using photons and trapped ions \cite{Podhora2022}. This framework offers the advantage of flexibility, enabling the construction of hierarchies that tolerate energy loss - a major challenge in many experiments aimed at detecting quantum non-Gaussian features. At its most stringent level, this framework has been theoretically formalized through the concept of stellar rank, specifically within the Fock-state basis \cite{Chabaud2020}.
 
This framework, however, does not address coherence, which is another fundamental resource in quantum physics \cite{Mandel1995,Streltsov2017}. Despite its importance, the understanding of this resource, especially in bosonic systems, remains an evolving area of research. In this context, we extend the classification framework to encompass quantum non-Gaussian (QNG) coherence. This approach has been experimentally tested with optical non-Gaussian states, as described in our companion paper \cite{Asenbeck2024}. 

Unlike the non-Gaussianity of Fock states, which focuses on diagonal elements of the density matrix, our framework isolates specific off-diagonal elements to eliminate interference from diagonal components. It serves as a complement to previously established resource theories of coherence, which rely on global mathematical measures that aggregate off-diagonal elements \cite{Baumgratz2014}. By directly targeting individual off-diagonal elements, this method provides detailed insights into quantum coherence that remain hidden in global measures. The individual coherences in bosonic systems exhibit significant diversity due to the varying strengths and sensitivities of Fock states. A hierarchical classification of these coherences leverages this diversity, offering significant advantages over traditional global measures. This framework enables a more precise and application-specific approach to understand and exploit quantum coherence in bosonic systems \cite{Baragiola2019,CampagneIbarcq2020}.

In this paper, we present a comprehensive study of  the proposed framework through a comparative analysis of two distinct hierarchies of criteria, each tailored to specific operational contexts and practical applications. We introduce a methodology for hierarchically classifying QNG coherence, offering a systematic way to interpreting and ranking the diverse range of individual coherences exhibited by bosonic systems. This classification is compared to contexts relevant to quantum sensing \cite{Giovannetti2006,McCormick2019} and error correction \cite{Michael2016}, where individual coherence elements play a critical role in enhancing performance and robustness. By exploring these hierarchies in detail, our analysis highlights how operational contexts influence the prioritization of specific coherence properties, offering deeper insights into the interplay between structure and function in quantum systems and its robustness to losses. This expanded perspective not only highlights the versatility of QNG coherence as a resource but also paves the way for refining its application-specific utility across a broad spectrum of quantum technologies.

The paper is organized as follows. Section \ref{SecII} introduces the framework for certifying specific quantum coherences by comparing them to the coherence limits achievable through Gaussian processes. Section \ref{SecIII} builds upon this framework by defining two distinct hierarchical certifications of quantum coherence, each tailored to a specific objective. Section \ref{SecIV} then presents a methodology for adapting these hierarchical certification criteria, enabling the evaluation of quantum coherence under less stringent experimental requirements. Finally, Sec. \ref{SecV} summarizes the results and provides an outlook.

%%%%%%%%%      
\section{Quantum non-Gaussian coherence}\label{SecII}
In this section, we introduce the framework for certifying non-Gaussian quantum coherences. We first define a suitable measure of coherence that captures its operational significance. We then quantify the coherence by comparing it to the maximum coherence achievable through classical or Gaussian processes.

\subsection{Quantum coherence measurement} \label{SecII.A}
\begin{figure}[t!]
	\includegraphics[width=0.9\columnwidth]{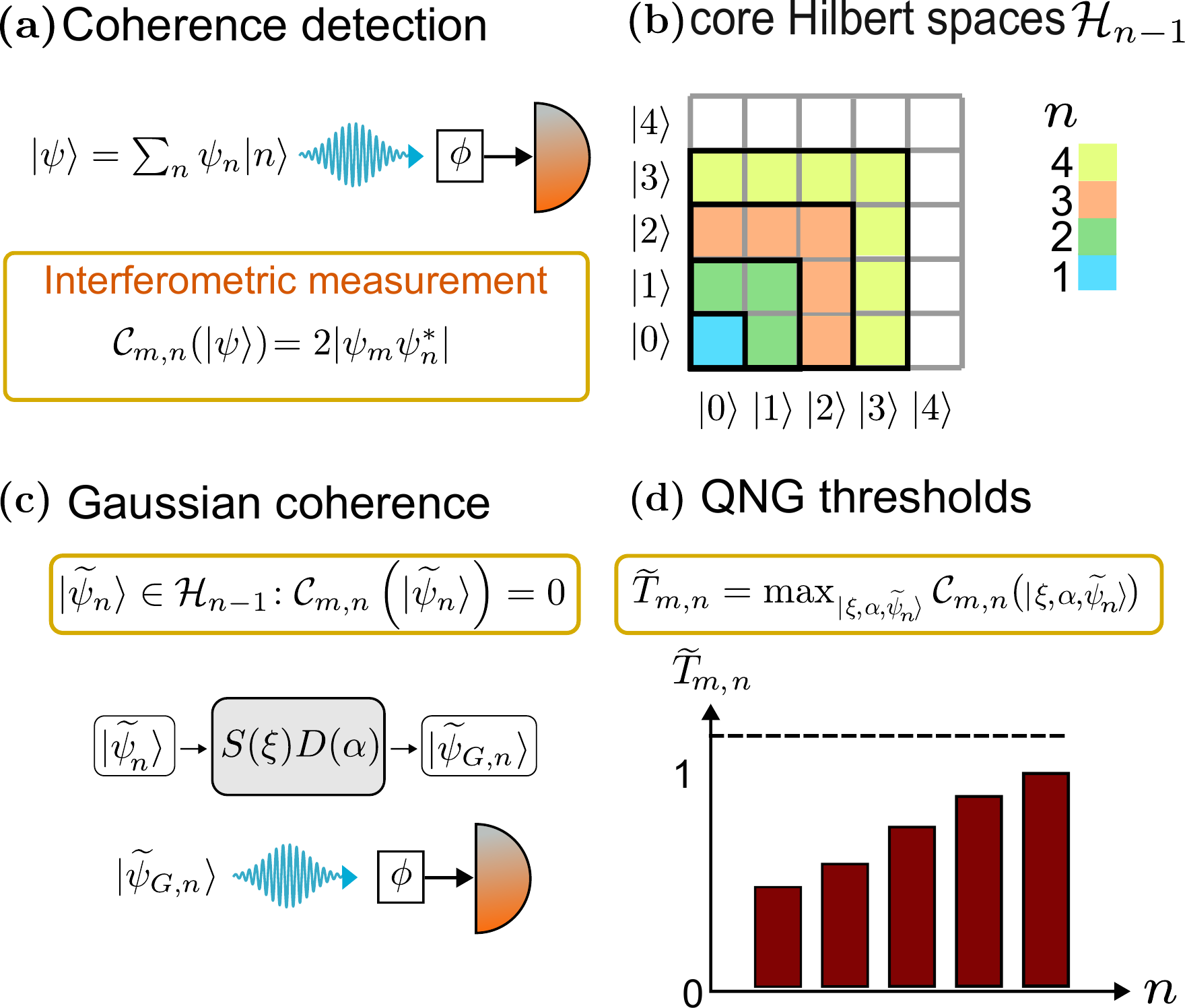}
	\caption{Introduction of non-Gaussian coherence. (a) For a pure state $|\psi\rangle=\sum_{n=0}^{\infty}\psi_n |n\rangle$, the coherence measure is given by $\mathcal{C}_{m,n}(|\psi\rangle)=2|\psi_m \psi_n^*|$. Detection of $\mathcal{C}_{m,n}$ relies on measurement of the phase-dependent projector $X_{m,n}(\phi)=\exp(i\phi)|m\rangle \langle n|+\exp(-i\phi)|n\rangle \langle m|$ and scanning its value over the phase $\phi$. (b) For each $\mathcal{C}_{m,n}$ with $m<n$, we introduce core Hilbert space $\mathcal{H}_{n}$ that the Fock states $|0\rangle$, ..., $|n-1\rangle$ span, as illustrated by the colored region in the grid. (c) A core state $|\widetilde{\psi}_n\rangle \in \mathcal{H}_{n}$ yields $\mathcal{C}_{m,n}(|\widetilde{\psi}_n\rangle)=0$. We define the non-Gaussian coherence as a value of $\mathcal{C}_{m,n}$ that we cannot achieve from the state $|\xi,\alpha,\widetilde{\psi}_{n}\rangle\equiv S(\xi)D(\alpha)|\widetilde{\psi}_n\rangle$, which corresponds to a core state $|\widetilde{\psi}_n\rangle$ affected by the squeezing operator $S(\xi)$ and the displacement operator $D(\alpha)$, which are the two Gaussian operations. (d) Certification of the QNG coherence imposes the condition $\mathcal{C}_{m,n}>\widetilde{T}_{m,n}=\max_{|\xi,\alpha,\widetilde{\psi}_{n}\rangle}\mathcal{C}_{m,n}(|\xi,\alpha,\widetilde{\psi}_{n}\rangle)$ for a particular coherence measure $\mathcal{C}_{m,n}$.}\label{fig:stellar}
\end{figure}

We investigate quantum coherence in the Fock-state basis and consider a Bloch sphere with the Fock states $|m\rangle$ and $|n\rangle$ at its poles. We focus on the coherence between the states located at the equator of the sphere. For a given state $\rho$, we quantify the targeted coherence by introducing the experimentally accessible coherence measure $\mathcal{C}_{m,n}(\rho)$ as
\begin{equation}\label{def:Cmn}
	\begin{aligned}
		\mathcal{C}_{m,n}(\rho)\equiv \frac{1}{2}\left\{\max_{\phi}\mbox{Tr}\left[\rho X_{m,n}(\phi)\right]\right.\\
		\left.-\min_{\phi}\mbox{Tr}\left[\rho X_{m,n}(\phi)\right]\right\},	
	\end{aligned}
\end{equation}
where $X_{m,n}(\phi)\equiv |m\rangle \langle n|\exp(i\phi)+|n\rangle\langle m|\exp(-i\phi)$ with $m\neq n$ corresponds to a phase-dependent projector. The definition in Eq.~(\ref{def:Cmn}) guarantees that $\mathcal{C}_{m,n}(\rho)$ is only sensitive to the off-diagonal element $|\langle m|\rho|n\rangle|$ irrespective of its phase, as depicted in Fig.~\ref{fig:stellar}(a). This definition allows $\mathcal{C}_{m,n}(\rho)$ to locally inspect the individual quantum coherences without mixing them with effects related to the diagonal elements. Equation~(\ref{def:Cmn}) directly follows interferometric Ramsey-type measurements \cite{McCormick2019} and does not, in principle, require full quantum state tomography.

\subsection{Coherence certification}\label{SecII.B}
The coherence measure $\mathcal{C}_{m,n}$ reaches its physical maximum $\mathcal{C}_{m,n}=1$ for any state of the ideal form $|\psi_{m,n}\rangle~\propto~|m\rangle+\exp(i\theta)|n\rangle$. Testing this coherence on a realistic state $\rho$ results in $\mathcal{C}_{m,n}(\rho)<1$. To test the achieved coherence, a first approach is to compare $\mathcal{C}_{m,n}(\rho)$ with the coherence reachable by any Gaussian state $|\xi,\alpha\rangle = S(\xi)D(\alpha)|0\rangle$ with $S(\xi)=\exp\left[\xi\left(a^{\dagger}\right)^2-\xi^* a^2\right]$ and $D(\alpha)=\exp\left(\alpha a^{\dagger}-\alpha^* a \right)$ being the squeezing and displacement operator, respectively (Fig. \ref{fig:stellar} (c)). Because the Gaussian states never saturate $\mathcal{C}_{m,n}$, we require
\begin{equation}\label{coherenceQNG}
	\mathcal{C}_{m,n}(\rho)>\widetilde{t}_{m,n}=\max_{\xi,\alpha}\mathcal{C}_{m,n}(|\xi,\alpha\rangle),
\end{equation}
to certify the quantum non-Gaussianity of $\rho$ based only on the measurement of the targeted coherence. As basis states, no Fock state can obey this criterion despite their non-Gaussian nature \cite{Lachman2019}. 

Since Fock states inherently exhibit no coherence we can extend the set of states that are rejected by the certification and calculate a new threshold. Similar to the resource theory of coherence \cite{Streltsov2017}, we consider any Fock state as 
a free state and the coherence arising from superpositions of Fock states as a resource. Incoherent mixing of Fock states is regarded as a free operation that does not yield coherence. In our framework, we also incorporate the Gaussian evolution $S(\xi)D(\alpha)$ into the free operations. Consequently, we also designate the coherence achieved by any state $|\xi,\alpha,n\rangle=S(\xi)D(\alpha)|n\rangle$ as free and consider that the QNG coherence requires $|\psi\rangle \neq |\xi,\alpha,n\rangle$ for any Fock state $|n\rangle$. To exclude coherence provided by these free states, we introduce the threshold
\begin{equation}\label{thresNonGC}
	t_{m,n}=\max_{\xi,\alpha,k}\mathcal{C}_{m,n}(|\xi,\alpha,k\rangle),
\end{equation}
which obeys $t_{m,n}\geq \widetilde{t}_{m,n}$ since $t_{m,n}$ covers a larger set of states. As an example, the numerically derived thresholds $t_{0,n}$ are $(0.93,0.71,0.63,0.55)$ for $n=\{1,...,4\}$ while thresholds $\widetilde{t}_{0,n}$ read $(0.93,0.71,0.50,0.46)$ for $n$ in the same range. The thresholds $t_{0,n}$ exceed the thresholds $\widetilde{t}_{0,n}$  for $n=3$ and $n=4$ as Gaussian evolution of Fock states $|1\rangle$ and $|2\rangle$ can overcome the coherence achieved only from Gaussian evolution of vacuum. Let us note that the thresholds decrease with $n$ as, for large $n$, Gaussian evolution acting on Fock state or vacuum increasingly spreads the coherence among many Fock states rather than focuses it on the targeted superposition. We numerically verified this tendency up to $n=10$, while keeping the set of rejected states constant. The threshold $t_{m,n}$ represents the most straight-forward definition of QNG coherences, by excluding any states that exhibit no coherence. However, we can develop  more selective thresholds by rejecting various Fock-state superpositions instead of only the Fock states and order them, which leads to hierarchies.

%%%%%%%%%%%
\section{Hierarchies of QNG coherence}\label{SecIII}
Expanding on the concept of more selective thresholds, we develop hierarchies for quantum non-Gaussian coherence. By rejecting various Fock-state superpositions instead of only the individual Fock states, we define a structured classification of coherence properties tailored to specific objectives. These hierarchies not only refine the certification process but also provide a framework to classify different forms of coherence based on their operational relevance.

\subsection{Certification and core states}

We refine the certification of targeted quantum non-Gaussian coherence by extending the set of states that the certification rejects. Let $\mathcal{H}_N$ denote the Hilbert space spanned by the Fock states $|n\rangle$ with $n\leq N$. We call a pure state $|\psi_N\rangle \in \mathcal{H}_N$ a \textit{core} state with respect to a given coherence measure $\mathcal{C}_{m,n}$ if $\mathcal{C}_{m,n}(|\psi_N\rangle)=0$. For a specific $\mathcal{C}_{m,n}$ with $m<n$, we can define two sets of core states, the first being $|\psi_{n,N}\rangle=\sum_{k\neq n}^{N}c_k |k\rangle$ and the second being $|\psi_{m,N}\rangle=\sum_{k\neq m}^{N}c_k |k\rangle$. Focusing on the first set, we then determine the maximal coherence measure achievable when only Gaussian dynamics act on $|\psi_{n,N}\rangle$, \textit{i.e.},
\begin{equation}\label{stellarT}
	T_{m,n}^{(N)}=\max_{\xi,\alpha,|\psi_{n,N}\rangle}\mathcal{C}_{m,n}\left(|\xi,\alpha,\psi_{n,N}\rangle\right),
\end{equation}
where $|\xi,\alpha,\psi_{n,N}\rangle=S(\xi)D(\alpha)|\psi_{n,N}\rangle$. For $(m,n)=(0,4)$ and $(m,n)=(3,4)$, we verified numerically that $T_{m,n}^{(N)}$ grows with $N$ and achieves $T_{m,n}^{(N)} \rightarrow 1$ for large $N$. 

To study if this convergence behavior is generic, we numerically derived $\log_{10}( 1-T_{0,n}^{(N)})$ and $\log_{10}( 1-T_{n-1,n}^{(N)})$ for various $n$ and observed significant dropping of both quantities between the cases with $N=10$ and $N=20$ (see Appendix). We repeated this analysis analogously with the second set of core states $|\psi_{m,N}\rangle=\sum_{k\neq m}^{N}c_k |k\rangle$ where $m$ is the lower index of the evaluated coherence measure $C_{m,n}$ and observed the respective thresholds exhibits $T_{m,n}^{(N)} \rightarrow 1$ similarly. The details of these calculations are provided in the Appendix \ref{AppendixA}. This means that Gaussian dynamics brings $\mathcal{C}_{m,n}$ very close to unity, if we allow for core states that exhibit any coherence except for the targeted ones and set no upper limit on $N$. As such, employing the thresholds $T_{m,n}^{(N)}$ with $N\gg 1$ for benchmarking implies the condition $\mathcal{C}_{m,n}=1$, which is unreachable for realistic states.

To obtain realistic experimental thresholds, we specify core states $|\widetilde{\psi}\rangle \in \mathcal{H}_N$ that are stricter compared to considering only the requirement $\mathcal{C}_{m,n}(|\widetilde{\psi}\rangle)=0$. We first impose an additional condition $|\widetilde{\psi}_n\rangle \in \mathcal{H}_{n-1} \subset \mathcal{H}_N$, where $\mathcal{H}_{n-1} $ is spanned by the Fock states $|k\rangle$ with $k<n$ (assuming $n\leq N$) and $n$ is given by the evaluated $\mathcal{C}_{m,n}$ with $m<n$, as depicted in Fig.~\ref{fig:stellar} (b). Thus, certification relies on beating the thresholds
\begin{equation}\label{stellarHTh}
	\widetilde{T}_{m,n}=\max_{\xi,\alpha,|\widetilde{\psi}_n\rangle\in \mathcal{H}_{n-1}}\mathcal{C}_{m,n}(|\xi,\alpha,\widetilde{\psi}_n\rangle),
\end{equation}
where $|\xi,\alpha,\widetilde{\psi}_n\rangle =S(\xi)D(\alpha)| \widetilde{\psi}_n\rangle$ and $|\widetilde{\psi}_n\rangle=\sum_{k=0}^{n-1}c_k|k\rangle$, showcased in Fig. \ref{fig:stellar} (d). While the maximal Fock state contributing to the state $|\widetilde{\psi}_n\rangle$ is specified by $n$ of the considered $\mathcal{C}_{m,n}$, the states $|\psi_{m,N}\rangle$ and $|\psi_{n,N}\rangle$ are restricted by missing contribution of either Fock state $|m\rangle$ or $|n\rangle$. Imposing the condition $\mathcal{C}_{m,n}>\widetilde{T}_{m,n}$ corresponds to traditional evaluation of the stellar rank \cite{Lachman2019,Chabaud2020} based on taking into account only the coherence measure. However, this is a more restrictive definition of core states than the free states covered by the thresholds in Eq.~(\ref{thresNonGC}), where we allowed for Gaussian evolution of the Fock states.

To maintain the certification based on Eq.~(\ref{thresNonGC}), we introduce a set of core states that includes all Fock states \( |n\rangle \in \mathcal{H}_N \) within a Hilbert space limited by a large \( N \). At the same time, these core states must be sufficiently restricted to avoid imposing the condition \( \mathcal{C}_{m,n} = 1 \) in the limit of large \( N \). This is achieved by defining a hierarchy of coherences.

\begin{figure*}[t!]
	\includegraphics[width=1.85\columnwidth]{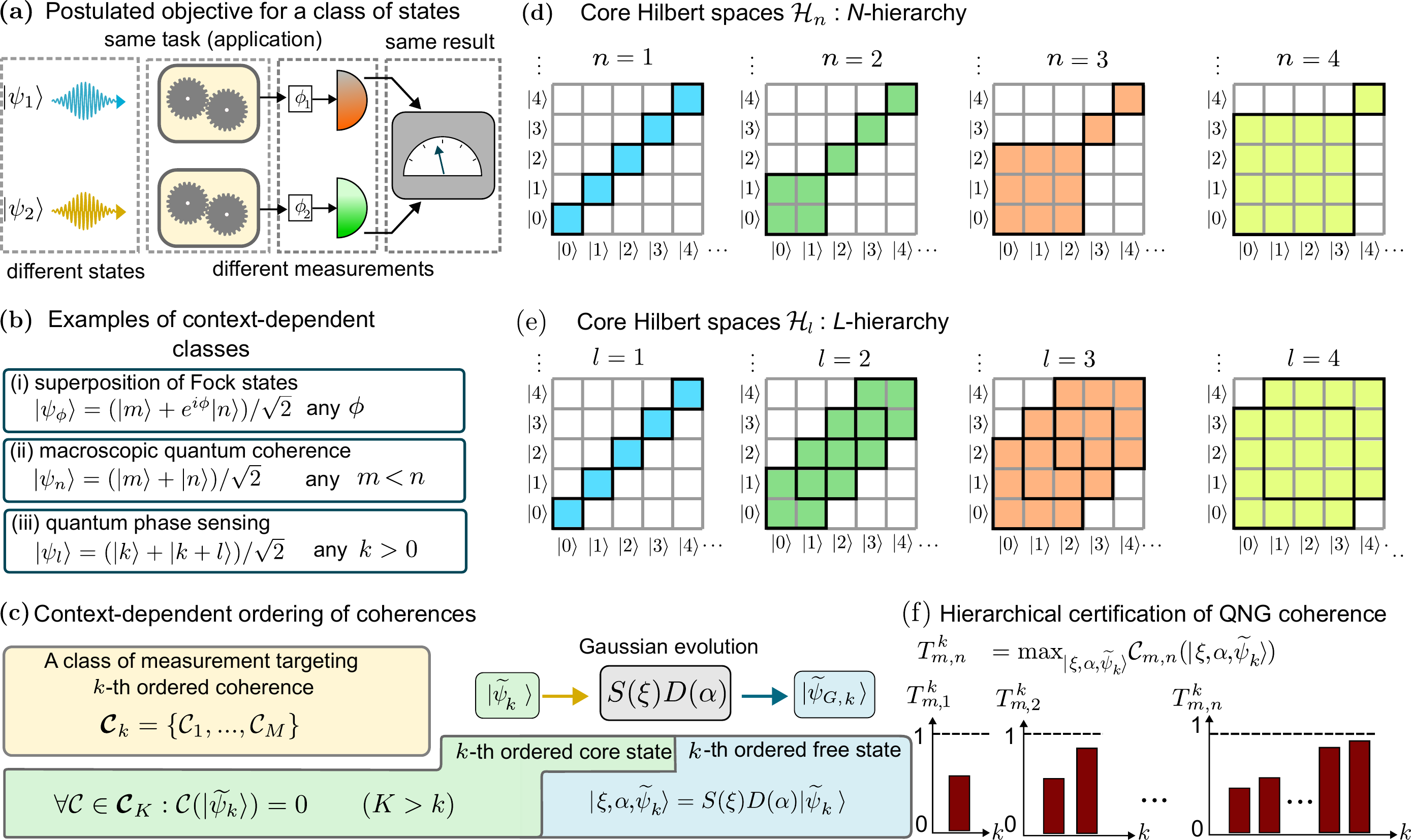}
	\caption{Context-dependent coherence framework and hierarchical certification (a) Motivation for a context-dependent class of states. We accomplish a specific task or application by employing the state $|\psi_1\rangle$ or the state $|\psi_2\rangle$, with coherence measures adjusted to each state $|\psi_1\rangle$ and $|\psi_2\rangle$.  Those outcomes have equivalent value for the considered task. (b) Examples of classes comprising states with an equal value with respect to a given objective. (i) The state $(|m\rangle+e^{i\phi}|n\rangle)/\sqrt{2}$ exhibit $\mathcal{C}_{m,n}=1$ regardless of the phase $\phi$. (ii) Macroscopic superposition $(|m\rangle+|n\rangle)/\sqrt{2}$ requires either $m$ or $n$ as high as possible. For example, $(|0\rangle+|n\rangle)/\sqrt{2}$ and $(|n-1\rangle+|n\rangle)/\sqrt{2}$ have the macroscopic coherence of the same order. (iii) In quantum sensing, the state $(|k\rangle+|k+l\rangle)/\sqrt{2}$ provides an advantage that scales with $l$, independent of $k$. (c) Set $\boldsymbol{\mathcal{C}}_k$ of different coherence measures. Each $\mathcal{C}\in \boldsymbol{\mathcal{C}}_k$ targets coherence with same value in an objective-oriented framework. Here, the index $k$ indicates postulated ordering of the coherence according to a chosen objective. We define a core state $|\widetilde{\psi}_k\rangle$ as a state that obeys $\mathcal{C}_K(|\widetilde{\psi}_k\rangle)=0$ for all $\mathcal{C}_K\in \boldsymbol{\mathcal{C}}_K$ with $K \geq k$. A free state $|\xi,\alpha,\widetilde{\psi}_{k}\rangle$ of order $k$ is obtained by applying Gaussian dynamics to a core state, \textit{i.e.}, $|\xi,\alpha,\widetilde{\psi}_{k}\rangle=S(\xi)D(\alpha)|\widetilde{\psi}_{k}\rangle$. (d) Introducting a $N$-hierarchy. This hierarchy prioritizes the targeted coherence $\mathcal{C}_{m,n}$ with high $n$, irrespective of $m<n$. Any core state $|\widetilde{\psi}_n\rangle$ with the order $n$ either belongs to the core Hilbert space $\mathcal{H}_{n}$, which Fock states $|k\rangle$ with $k<n$ span, or corresponds to an arbitrary Fock state as illustrated by the colored squares in the respective grid. (e) Introducing a $L$-hierarchy. This hierarchy prioritizes coherence targeted by $\mathcal{C}_{m,n}$ with a large difference $l = n-m$. A core state $|\widetilde{\psi}_l\rangle$ belongs to a core Hilbert space $\mathcal{H}_{l}$, which the Fock states $|k\rangle$, ..., $|k+l-1\rangle$ span (with arbitrary $k$), as depicted by colored squares in each grid. While these squares, representing the core Hilbert spaces, can be shifted along the diagonal direction, they always preserve their size, given by the order. (f) Coherence thresholds. For each coherence measure $\mathcal{C}_{m,n}\in \boldsymbol{\mathcal{C}}_n$, we derive a sequence of thresholds ordered by the index $k$. The $k^{\textrm{th}}$-threshold covers all $C_{m,n}$ that the free states $|\widetilde{\psi}_{G,k}\rangle$ reach. Because we allow for context-dependent ordering of free states, the thresholds depend on the considered hierarchy.}\label{fig:Hierarchies} 
\end{figure*}

Given target states of the form $(1/\sqrt{2})(\ket{m}+\ket{n})$, we can define an ordering property $k$, which is maximized by one or more target states, defined by $m$ and $n$. We quantify the coherence of states with the property $k$ by $\mathcal{C}_{m(k),n(k)}$ where the subscript $m(k),n(k)$ means that the indexes $m$ and $n$ depend formally on the property $k$. Further, we employ the notation $\mathcal{C}_{k}$ instead of $\mathcal{C}_{m(k),n(k)}$. Assuming $M$ target states maximize the property $k$, we introduce the set \( \boldsymbol{\mathcal{C}}_{k} = \{\mathcal{C}_{1,k}, \dots, \mathcal{C}_{M,k}\} \). We can then consider that $k \in \{1,2,....,k_{\max}\}$ and allow for ordered sets $\boldsymbol{\mathcal{C}}_{1},\dots,\boldsymbol{\mathcal{C}}_{k_{\max}}$, which leads to the definition of a core state \( |\widetilde{\psi}_k\rangle \) of order \( k \) by requiring \( \mathcal{C}_{K}(|\widetilde{\psi}_k\rangle) = 0 \) for all \( \mathcal{C}_{K} \in \boldsymbol{\mathcal{C}}_{K} \) with \(K \geq k \), as illustrated in Fig.~\ref{fig:Hierarchies} (c).

Core states of a given ordering property \( k \) are not restricted to a single Hilbert space but belong to a core Hilbert space \( \mathcal{H}_m \) within a set \( \boldsymbol{\mathcal{H}} = \{\mathcal{H}_1, \dots, \mathcal{H}_M\} \). We use the notation \( |\widetilde{\psi}\rangle \in \boldsymbol{\mathcal{H}} \) to indicate that \( |\widetilde{\psi}\rangle \) resides in one of the core spaces \( \mathcal{H}_m \in \boldsymbol{\mathcal{H}} \).

A state is considered \textit{free} in a given order if it can be generated by Gaussian dynamics applied to a core state. The ordering of core states naturally induces an ordering of free states. Hierarchical certification targets coherence by outperforming only those free states that reach a specific order, as shown in Fig.~\ref{fig:Hierarchies} (f). This approach, depicted in  Fig.~\ref{fig:Hierarchies} (a) makes the evaluation context- or $k-$dependent, while confining the set of free states to avoid imposing the unattainable condition \( \mathcal{C}_{m,n} = 1 \), which would occur without ordering the coherence resources.

To demonstrate the proposed framework, we analyze two hierarchies based on distinct contexts as in  Fig.~\ref{fig:Hierarchies} (b) for evaluating coherence.

\subsection{Macroscopic quantum non-Gaussian coherence: Large Fock numbers}
We first consider an example of a rather fundamental objective: achieving macroscopic non-Gaussian coherence by favoring superpositions of high Fock states. Specifically, this objective targets states of the form $(|m\rangle+|n\rangle)/\sqrt{2}$ with the Fock state $|n\rangle$ being as high as possible, irrespective of the Fock state $|m\rangle$ (assuming $m<n$). Such coherence has already been explored experimentally \cite{Turchette2000,Myatt2000,Wang2017,Lu2021,Chu2018}. To formalize this, we specify that any measurement $\mathcal{C}_{m,n}$ with $m<n$ targets coherence with the same hierarchical order $n$. 

Let $\boldsymbol{\mathcal{H}}_{n}$ denote the set of core states $|\widetilde{\psi}_{n}\rangle$ that obey $\mathcal{C}_{m,n}(|\widetilde{\psi}_n\rangle)=0$ (with $m<n$) for all $m<n$. Any Fock state $|k\rangle$ is a core state since $\mathcal{C}_{m,n}(|k\rangle)=0$. Also, the superposition $|\widetilde{\psi}_{n}\rangle=\sum_{k=0}^{n-1}c_k|k\rangle$ reaches $\mathcal{C}_{m,n}= 0$ (irrespective to $m<n$), and therefore $|\widetilde{\psi}_{n}\rangle \in \mathcal{H}_{n}$. We certify the quantum non-Gaussian coherence in the $k-$th order if the coherence measure $\mathcal{C}_{m,n}$ exceeds the threshold
\begin{equation}\label{NHierachy}
	T_{m,n}^{N,k}=\max_{\xi,\alpha,|\widetilde{\psi}_k\rangle\in\boldsymbol{\mathcal{H}}_{k}}\mathcal{C}_{m,n}(|\xi,\alpha,\widetilde{\psi}_{k}\rangle),
\end{equation}
where $|\xi,\alpha,\widetilde{\psi}_k\rangle=S(\xi)D(\alpha)|\widetilde{\psi}_k\rangle$ corresponds to a free state in the $k$-th order. Note that only $k < n$ implies a beatable threshold $T_{m,n}^{N,k}<1$. We call the certification in Eq.~(\ref{NHierachy}), aiming at superposition of high Fock states, the \textit{$N$-hierarchy} of quantum non-Gaussian coherence.

\subsection{Quantum non-Gaussian coherence for sensing or error correction: Fock separation}

Another possibility is to order the coherence measure $\mathcal{C}_{m,n}$ with respect to the difference $l=n-m$, which has a significant role in leveraging coherence for specific bosonic applications, like quantum sensing \cite{Giovannetti2006} and error correction using binomial codes \cite{Michael2016}. Accordingly, we state that the index $l$ in $\mathcal{C}_{k,k+l}$ corresponds to the hierarchical order irrespective of the index $k$. We specify a set of core states $\boldsymbol{\mathcal{H}}_{l}$ for such ordering of the coherence. Due to the requirement $\mathcal{C}_{k,k+l}(|\widetilde{\psi}_l\rangle)=0$ for all $k$, we find out that $\boldsymbol{\mathcal{H}}_{l}$ involves the superposition $|\widetilde{\psi}_{l}\rangle=\sum_{m=k}^{k+l-1}c_m|m\rangle$ with any $k$. In this context, non-Gaussian coherence in the order $l$ is achieved when the coherence measure surpasses the threshold
\begin{equation}\label{LHierachy}
	T_{k,l}^{L,r}=\max_{\xi,\alpha,|\widetilde{\psi}_r\rangle\in\boldsymbol{\mathcal{H}}_{r}}\mathcal{C}_{k,k+l}(|\xi,\alpha,\widetilde{\psi}_r\rangle),
\end{equation}
where $r< l$ and $|\xi,\alpha,\widetilde{\psi}_r\rangle=S(\xi)D(\alpha)|\widetilde{\psi}_r\rangle$ represents a free state of the $r$-th order. We call this certification, which favors the distance between two Fock states in the superposition, the \textit{$L$-hierarchy}.

\subsection{Comparison of hierarchies}

The $N$-hierarchy and the $L$-hierarchy rely on two distinct orderings of the coherence in the Fock-state basis, as shown in Fig.~\ref{fig:Hierarchies} (d) and (f). Consequently, the thresholds associated to the measurement $\mathcal{C}_{m,n}$ depend on which hierarchy is being evaluated. In general, the $N$-hierarchy imposes stricter requirements than the $L$-hierarchy when we test the targeted coherence $\mathcal{C}_{m,n}$ for large values of both $m$ and $n$. Conversely, the $L$-hierarchy is more demanding in some orders when we target the coherence $\mathcal{C}_{0,n}$. 

To highlight these differences, Fig.~\ref{comparison} (a)-(d) compares the thresholds for the coherence measures $\mathcal{C}_{0,4}$ and $\mathcal{C}_{3,4}$. The figure shows that the highest thresholds for $\mathcal{C}_{0,4}$ are identical for both hierarchies, \textit{i.e.}, $T_{0,4}^{N,4}=T_{0,4}^{L,4}=0.8$. However, the second-order threshold of the $L$-hierarchy $T_{0,4}^{L,2}=0.62$ is stricter compared to the corresponding threshold of the $N$-hierarchy, which is $T_{0,4}^{N,2}=0.55$. For $\mathcal{C}_{3,4}$, the $L$-hierarchy allows testing only one threshold, $T_{3,1}^{L,1}=0.8$. In contrast, the $N$-hierarchy can be evaluated by the thresholds up to the third order, which gives $T_{3,4}^{N,4}=0.96$. This comparison illustrates the inherent ambiguity in hierarchical certifications of the quantum non-Gaussian coherences. The context-dependent ordering of the core states implies that the requirements of the criteria can vary according to the chosen context.
\subsection{Non-Gaussian coherence depth}
We now evaluate the experimental requirements for the $N$-hierarchy and $L$-hierarchy under the influence of realistic imperfections. To model such imperfections, we consider bosonic leakage and thermalization of an ideal state $|\psi_{m,n}\rangle=(|m\rangle+|n\rangle)/\sqrt{2}$. 

\begin{figure*}[t]
	\includegraphics[width=1.85\columnwidth]{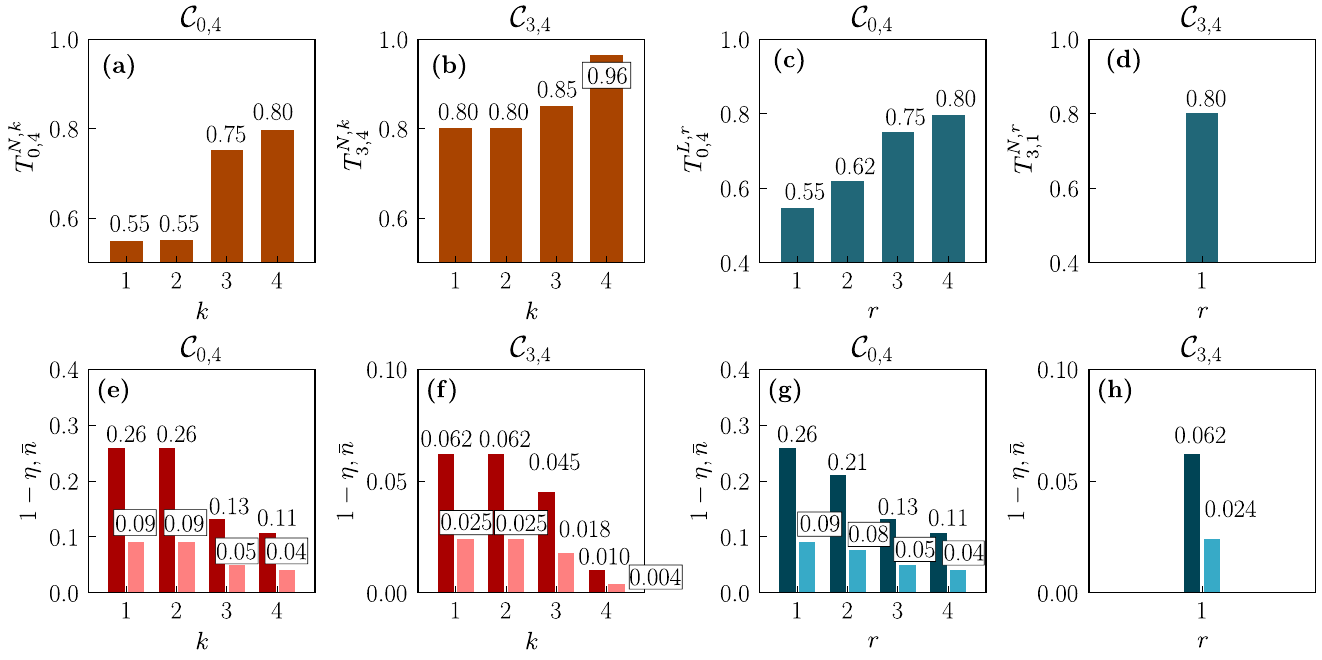}
	\caption{Absolute non-Gaussian coherence and criteria depth. (a)-(b) Absolute criteria in $N$-hierarchy. The absolute criteria for $\mathcal{C}_{0,4}$ in (a) and $\mathcal{C}_{3,4}$ in (b) require exceeding the thresholds $T_{0,4}^{N,k}$ and $T_{3,4}^{N,k}$, respectively. The horizontal axis represents the hierarchical order $n$, which determines the set $\boldsymbol{\mathcal{H}}_{n}$ of rejected core states. This set includes any Fock state and all the superposition $\sum_{k=0}^{n-1}c_k|k\rangle$. (c)-(d) Absolute criteria in the context of $L$-hierarchy. The absolute criteria for $\mathcal{C}_{0,4}$ in (c) and $\mathcal{C}_{3,4}$ in (d) require exceeding the thresholds $T_{0,4}^{L,k}$ and $T_{3,1}^{L,k}$, repectively. The index $l$ represents the hierarchical order that defines the set of rejected core states $\boldsymbol{\mathcal{H}}_{l}$ involving the superposition $\sum_{k=m}^{m+l-1}c_k|k\rangle$ with an arbitrary index $m$. We can compare $\mathcal{C}_{3,4}$ only with the threshold representing the first order. (e)-(h) Loss and thermal depth analysis. Each absolute criterion tolerates only a limited amount of loss (thermalization) affecting the ideal state $(|0\rangle+|4\rangle)/\sqrt{2}$ and $(|3\rangle+|4\rangle)/\sqrt{2}$, as modelled in Eq.~(\ref{model}). In (e) and (f), the dark red bars present the loss depth, defined as maximal probability $1-\eta$, while the light red bars represent the thermal depth, defined as the maximal mean number of noisy bosons $\bar{n}$, for which the threshold $T_{0,4}^{L,k}$ and $T_{3,4}^{l,k}$ can still be surpassed. Similarly, (g) and (h) depict the corresponding loss and thermal depth for the $L$-hierarchy. The numerical values attached to each bar indicate the respective depths achieved.}\label{comparison} 
\end{figure*}

Assuming that beating the thresholds tolerates only small imperfections, the model state $\rho(\eta,\bar{n})$ follows as \cite{Loudon2000}:
\begin{equation}\label{model}
	\begin{aligned}
		&\rho_{m,n}(\eta,\bar{n}) \propto |\psi\rangle\langle \psi|+\eta a |\psi\rangle\langle \psi|a^{\dagger}\\
		&+\bar{n}\left[a |\psi\rangle\langle \psi|a^{\dagger}+a^{\dagger}|\psi\rangle\langle \psi|a\right.\\
		&\left.-\left(a^{\dagger}a+1/2\right)|\psi\rangle\langle\psi|-|\psi\rangle\langle\psi|\left(a^{\dagger}a+1/2\right)\right],
	\end{aligned}
\end{equation}
where $\eta$ corresponds to the transmission and $\bar{n}$ quantifies thermalization of the ideal state. This model is valid only in the limit $1-\eta\ll 1$ and $\bar{n}\ll 1$. 

Based on the state $\rho_{m,n}(\eta,\bar{n})$, we introduce the loss depth as the minimal transmission $\eta$ that still allows beating a respective threshold by the coherence measure $\mathcal{C}_{m,n}(\eta)$. Similarly, we define the thermal depth as the maximal $\bar{n}$ that permits exceeding the threshold. Figure~\ref{comparison} (e)-(h) compares both depths in the context of the $N$-hierarchy and the $L$-hierarchy focusing on the coherence measures $\mathcal{C}_{0,4}$ and $\mathcal{C}_{3,4}$. A notable difference occurs for $\mathcal{C}_{3,4}$: the highest testable threshold of $N$-hierarchy implies both depth $1-\eta$ and $\bar{n}$ are seven times smaller compared to the highest testable threshold of the $L$-hierarchy. This illustrative example demonstrates various requirements that stem from the context-dependent certification of the quantum non-Gaussian coherence.

\section{Hierarchy of relative criteria}\label{SecIV}
In the previous sections, we developed hierarchies of coherence criteria that impose strict requirements based solely on the coherence measure. While these \textit{absolute} criteria provide a clear characterization of coherence, they may be too strict for certain applications. To address this limitation, we introduce a more flexible approach by incorporating additional constraints, specifically diagonal elements of the density matrix. We refer to these new criteria as \textit{relative} criteria. This section introduces and develops a family of relative criteria, exploring their use in both hierarchical structures.

\subsection{Incorporating additional constraints}

We introduce a family of relative criteria for any hierarchical order in both $N$- and $L$-hierarchy. Although relative criteria reject the same set of states as the absolute criteria, they may reduce requirements below that of absolute criteria and still form a hierarchy. We can allow for a \textit{two-dimensional} relative criterion combining the coherence measure $\mathcal{C}_{m,n}$ with any probability $P_k=\langle k|\rho|k\rangle$. However, since we aim at the coherence reached by states approaching a particular target state $(|m\rangle+|n\rangle)/\sqrt{2}$, we assume that only the probabilities $P_m$ and $P_n$ achieve significant values. Thus, we restrict the certification only to the criteria including the probability $P_m$, $P_n$ or the probability of multi-photon error $P_{e,n}=1-\sum_{k=0}^n P_k$. More generally, we can also extend the procedure by considering \textit{three-dimensional} relative criteria that combine $C_{m,n}$, $P_n$ and $P_{e,n}$. In the following, we describe a procedure for deriving these criteria and demonstrate them on some examples.

\begin{figure}[t]
	\includegraphics[width=0.9\columnwidth]{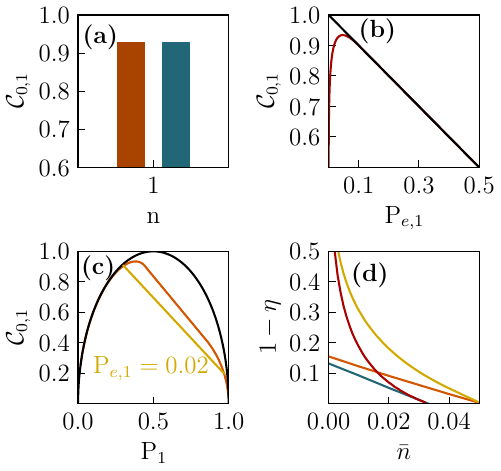}
	\caption{Relative criteria. (a) The absolute criterion for the coherence measure $\mathcal{C}_{0,1}$ is identical for both the $N$-hierarchy (red bar) and the $L$-hierarchy (blue bar) as the same set of states is rejected in this case. (b) The red line represents the threshold required to satisfy a relative criterion using the coherence measure $\mathcal{C}_{0,1}$ and the probability $P_{e,1}$ of two and more photons. The black line indicates the boundary for all physical states. (c) The red line corresponds to the threshold for $\mathcal{C}_{0,1}$ as a function of the probability $P_1$ of one photon. The yellow line presents the threshold taking also into account $P_{e,1}$, with $P_{e,1}=0.02$ as example. This demonstrates that the threshold decreases when additional constraints on $P_{e,1}$ are introduced. The black line depicts the physical boundary assuming no knowledge about the probability $P_{e,1}$. (d) The capability of a decohered state to beat the thresholds is illustrated. The model assumes that the ideal state $(|0\rangle+|1\rangle)/\sqrt{2}$ undergoes loss and thermalization quantified by the parameters $\eta$ and $\bar{n}$, as described by Eq.~(\ref{model}). The lines represent the upper boundary of the parameters $\bar{n}$ and $\eta$ for fulfilling various criteria: the absolute criterion (blue line), the criterion incorporating $P_{e,1}$ (red line), the criterion incorporating $P_1$ (orange line), and the criterion incorporating both $P_1$ and $P_{e,1}$ (yellow line).}\label{fig4:RelativeCriteria} 
\end{figure}

\begin{figure*}[ht]
	\includegraphics[width=1.8\columnwidth]{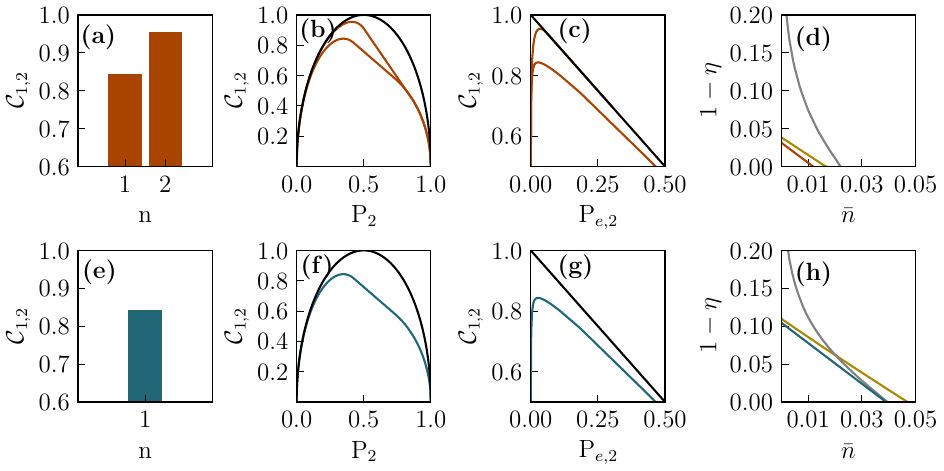}
	\caption{Analysis of the coherence $\mathcal{C}_{1,2}$. (a-d) Certification of $\mathcal{C}_{1,2}$ using the $N$-hierarchy.  The absolute criterion in (a) requires $\mathcal{C}_{1,2}>0.84$ for testing the first hierarchical order and $\mathcal{C}_{1,2}>0.96$ for testing the second order. The two red lines in (b) and (c) represent the thresholds for relative criteria depending on the probability $P_2$ of two photons and on the probability $P_{e,2}$ of three and more photons, respectively. The panel (d) provides the maximal loss and thermalization affecting an ideal state $(|1\rangle+|2\rangle)/\sqrt{2}$ required to exceed the absolute second-order threshold. States with parameters below the red line can be certified. The yellow and gray line indicate the thresholds for states that obey the relative second-order criterion involving $P_2$ and $P_{e,2}$, respectively. (e-h) Certification of $\mathcal{C}_{1,2}$ using the $L$-hierarchy. Only the first hierarchical order can be tested. The absolute criterion in (e) requires $\mathcal{C}_{1,2}>0.84$. The blue lines in (f) and (g) represent the thresholds for the relative criterion depending on the probability $P_2$ and the probability $P_{e,2}$, respectively. The panel (h) provides the loss and thermalization required to exceed the absolute criterion. The yellow line and the gray line indicate the threshold for states that obey the relative first-order criterion involving $P_2$ and $P_{e,2}$, respectively. The black lines represent the physical boundary.}\label{fig5:RelativeCriteria} 
\end{figure*}

\subsection{Two-dimensional relative criteria}

Two-dimensional criteria impose a condition on the pair $(P,\mathcal{C}_{m,n})$ with $P \in \{P_m,P_n,P_{e,n}\}$. To derive a criterion, we allow for the convex linear combination \cite{Filip2011}
\begin{equation}
	F_{\lambda,m,n}(\rho)=\mathcal{C}_{m,n}(\rho)+\lambda P(\rho),
\end{equation}
where $\lambda$ is a free parameter. Then, we need to calculate the maximum 
\begin{equation}
F_{m,n}(\lambda)=\max_{\xi,\alpha,|\widetilde{\psi}_{k}\rangle \in \boldsymbol{\mathcal{H}}_k} F_{\lambda,m,n}(|\xi,\alpha,\widetilde{\psi}_{k}\rangle)
\end{equation}
over all the states $|\xi,\alpha,\widetilde{\psi}_{k}\rangle=S(\xi)D(\alpha)|\widetilde{\psi}_{k}\rangle$ where $\boldsymbol{\mathcal{H}}_k$ is a set of core states according to the tested hierarchy. First, we use the procedure in Ref. \cite{Fiurasek2022} to maximize over a given core state $|\widetilde{\psi}_{l}\rangle \in \mathcal{H}_l$, where $\mathcal{H}_l$ is a particular Hilbert space identified by the index $l$. It results in the function
\begin{equation}
	\widetilde{F}_{\lambda,m,n}(\xi,\alpha,\phi,l,\theta)=\max_{|\widetilde{\psi}_{l}\rangle}F_{\lambda,m,n}(|\xi,\alpha,\widetilde{\psi}_{l}\rangle).
\end{equation}
A threshold corresponds to a global maximum of the function $\widetilde{F}_{\lambda,m,n}(\xi,\alpha,\phi,l,\theta)$. 

Maximizing this function is technically difficult due to a large number of local maxima. Since using global optimizing methods is typically slow and does not converge well, we simplify the task. First, we guess for which parameters the maximum occurs and, then, we use random sampling of the parameters $\xi$, $\alpha$, $\phi$, $l$ and $\theta$ to verify the estimated maximum for different $\lambda$. 
Specifically, we make a conjecture that the maximum occurs for $\phi=0$, $\theta \in \{0,\pi\}$ and $|\widetilde{\psi}_{l}\rangle \in \{\sum_{k=0}^{l-1}c_k|k\rangle,\sum_{k=1}^{l}c_k|k\rangle\}$ ($L$-hierarchy) or $|\widetilde{\psi}_{l}\rangle \in \{\sum_{k=0}^{l-1}c_k|k\rangle,|l\rangle\}$ ($N$-hierarchy). Assuming this, we use the equation
\begin{equation}\label{SM:alfEq}
	\partial_{\alpha}\widetilde{F}_{\lambda,m,n}(\xi,\alpha,0,l,\theta)=0
\end{equation}
to determine $\alpha$ maximizing $\widetilde{F}_{\lambda,m,n,l}(\xi,\alpha,0,l,\theta)$. Due to rules for the differentiation of analytical functions and an algebraic procedure employed to derive eigenvalues of a matrix \cite{Fiurasek2022}, we can reformulate Eq.~(\ref{SM:alfEq}) as a polynomial equation of the variable $\alpha$ and obtain numerically the polynomial roots for any given $\xi$, $l$ and $\theta$. Then, the function $H_{\lambda,m,n}(\xi,l,\theta)\equiv \max_{\alpha}\widetilde{F}_{\lambda,m,n}(\xi,\alpha,0,l,\theta)$ can be maximized over $\xi$ and the index $l$, which leads to the maximum $F_{m,n}(\lambda)$, depending only on the parameter $\lambda$. A criterion follows as:
\begin{equation}\label{2DCrit}
	C_{m,n}(P)>\min_{\lambda} \left[F_{m,n}(\lambda)-\lambda P(\rho)\right],
\end{equation}
where $P$ is a measured probability that depends on the tested state $\rho$.
Thus, it amounts to determine $\lambda$ minimizing the right-hand side of Eq. \ref{2DCrit}.

\subsection{Three-dimensional relative criteria}

The requirement of a criterion combining $\mathcal{C}_{m,n}$ with either $P_n$ or $P_{e,n}$ can be further relaxed by incorporating both probabilities $P_n$ and $P_{e,n}$ simultaneously. To derive such a criterion, we consider the convex linear combination
\begin{equation}
	F_{\lambda_1,\lambda_2,m,n}(\rho)=\mathcal{C}_{m,n}(\rho)+\lambda_1 P_n(\rho)+\lambda_2 P_{e,n}(\rho)
\end{equation}
and derive the maximum 
\begin{equation}
F_{m,n}(\lambda_1,\lambda_2)=\max_{\xi,\alpha,|\widetilde{\psi}_{l}\rangle\in \mathcal{H}_k}F_{m,n,\lambda_1,\lambda_2}(|\xi,\alpha,\widetilde{\psi}_{l}\rangle).
\end{equation}
Analogous to the derivation of two-dimensional criteria, we formulate the criterion as
\begin{equation}\label{3DCrit}
	\mathcal{C}_{m,n}(P_m,P_{e,n})>\min_{\lambda_1,\lambda_2}\left[F_{m,n}(\lambda_1,\lambda_2)-\lambda_1 P_m-\lambda_2 P_{e,n}\right].
\end{equation}
Minimizing the right-hand side of Eq.~\ref{3DCrit} has to be done numerically. Such derivation for an example with $\mathcal{C}_{0,1}$, $P_1$ and $P_{e,1}$ is detailed in Appendix \ref{AppendixB}.

\subsection{Comparative analysis of the criteria}

To illustrate the multi-dimensional relative criteria, we provide two figures. Figure \ref{fig4:RelativeCriteria} presents the thresholds for $\mathcal{C}_{0,1}$ taking into account $P_0$, $P_1$ and $P_{e,1}$, while Fig. \ref{fig5:RelativeCriteria} details the thresholds for $\mathcal{C}_{1,2}$ with $P_2$ and $P_{e,2}$. Both show how the thresholds are modified by incorporating additional constraints.

Specifically, Fig.~\ref{fig4:RelativeCriteria} illustrates the procedure of introducing the relative criteria for testing the coherence measure $\mathcal{C}_{0,1}$. The absolute criterion shown in Fig.~\ref{fig4:RelativeCriteria} (a) imposes $\mathcal{C}_{0,1}>0.93$. We lower this requirement by incorporating the probability $P_{e,1}$, the probability $P_1$ or both $P_{e,1}$ and $P_1$ into the tested criterion, as shown in Fig.~\ref{fig4:RelativeCriteria} (b)-(d).  Figure ~\ref{fig5:RelativeCriteria} depicts the relative criteria derived for $\mathcal{C}_{1,2}$ in the context of both hierarchies. The $N$-hierarchy in Fig. ~\ref{fig5:RelativeCriteria} (a)-(d) allows for testing first-order relative criteria and second-order relative criteria, the $L$-hierarchy in ~\ref{fig5:RelativeCriteria} (e)-(h) enables only the evaluation of the first-order relative criteria. It demonstrates that the context-dependent certification, introduced for the absolute criteria, can be transferred to the relative criteria.

In both figures, the capability of a decohered state to beat the thresholds is also presented. The absolute criteria impose the strictest condition on the transmission $\eta$ and the noise $\bar{n}$. Figures ~\ref{fig4:RelativeCriteria} and \ref{fig5:RelativeCriteria} show how this condition can be lowered. By including the probability $P_{e,1}$ or $P_{e,2}$, the criterion tolerates arbitrarily large losses if thermal noise does not deteriorate a tested state. However, realistic states always exhibit some noise contributions, which implies that the criterion with $P_{e,1}$ or $P_{e,2}$ imposes a condition on $\eta$, depending on $\bar{n}$.

\section{Conclusion}\label{SecV}
Compared to the resource theory of quantum coherence, our methodology offers a more detailed and hierarchical non-Gaussian quantum analysis of bosonic quantum coherence, focusing solely on individual off-diagonal elements in the Fock-state basis. It is directly applicable to a wide range of experiments in quantum photonics \cite{Ourjoumtsev2006,Huang2015,Cotte2022,Darras2022,Konno2024}, trapped ions \cite{Harty2014,McCormick2019,Fluehmann2019}, neutral atoms \cite{Magro2023} and superconducting circuits \cite{Wang2017,Wang2019,Lu2021}, where such coherences are experimentally investigated and used for applications in quantum sensing, quantum computing and fundamental studies in quantum thermodynamics. This hierarchical framework allows for experiments to certify whether the quantum non-Gaussian coherences can drive particular applications or fundamental tests. With the examples of the two hierarchies, we demonstrated a diversity of hierarchical classifications of quantum non-Gaussian coherences, depending on the context of evaluation. Additionally, we introduced a methodology for constructing more tolerant hierarchies of relative criteria, which are particularly useful when experimental conditions are challenging while still allowing complete hierarchical comparisons.  

When multiple individual coherences simultaneously play a crucial role in fundamental tests or applications, the methodology can be further extended to analyse coherence of unbalanced superpositions of Fock states or trade-offs between different quantum non-Gaussian coherences, similarly to what has been started for nonclassical Fock probabilities \cite{Innocenti2022} and coherences \cite{Innocenti2023}. Additionally, the framework can be further developed to address two-mode and eventually multi-mode quantum non-Gaussian coherences, that can be used for dual-rail binomial codes and in sensing protocols, following the previous works on quantum non-Gaussian coincidences \cite{Lachman2021,Liu2024}. As the derivation of the thresholds become for such complex cases much more numerical, optimization methods will be necessary to derive the thresholds required for conclusive classification of fundamental tests and applications. This classification is essential for further experimental developments of quantum non-Gaussian resources for quantum technology \cite{Mattia} as demonstrated repeatedly in earlier experimental tests of criteria developed for diagonal elements \cite{Jezek2012,Straka2014,Straka2018,Lachman2019}.

\section*{Acknowledgements}
This work was supported by the French National Research Agency via the France 2030 project Oqulus (ANR-22-PETQ-0013). L.L. acknowledges the support from project No. 23-06015O and R.F from project 23-06308S, both of the Czech Science Foundation. R.F. also acknowledges funding from the MEYS of the Czech Republic (Grant Agreement 8C22001) and CZ.02.01.01/00/22\_008/0004649 (QUEENTEC). This work was also jointly supported by the SPARQL and CLUSTEC projects that have received funding from the European Union's Horizon 2020 Research and Innovation Programme under Grant Agreement No. 731473 and 101017733 (QuantERA), and No. 101080173 (CLUSTEC). J.L. is a member of the Institut Universitaire de France. 

\begin{appendix}
\section{Convergence of the thresholds with increasing core Hibert space}\label{AppendixA}
\begin{figure}[t!]
	\includegraphics[width=0.8\columnwidth]{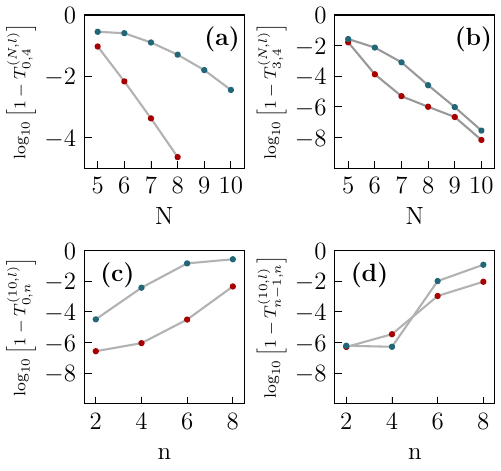}
	\caption{Convergence of the thresholds $T_{m,n}^{(N,l)}$ (with $m<n$) covering all values of coherence measure $C_{m,n}$ that we can reach by the Gaussian dynamics of core states $|\widetilde{\psi}_{l,N}\rangle=\sum_{k\neq l}^N c_k|k\rangle$ with either $l=m$ (blue points) or $l=n$ (red points). We demonstrate how tightly the thresholds approach one. (a) and (b) depict $1-T_{0,4}^{(N,l)}$ and $1-T_{3,4}^{(N,l)}$ in logarithmic scale. Whereas the blue points depict thresholds covering core states without the vacuum in (a) or the Fock $|3\rangle$ in (b), the red points show thresholds that cover core states without the Fock state $|4\rangle$. (c) and (d) illustrate the thresholds with various $n$ for $N=10$ with either $l=m$ (blue points) or $l=n$ (red points)..}\label{fig:SMConvergency} 
\end{figure}

We allow for the core states $|\psi_{N,m}\rangle=\sum_{k\neq m}^{N}c_k |k\rangle$ and $|\psi_{N,n}\rangle=\sum_{k\neq n}^{N}c_k |k\rangle$, where $m<n$. For a given $N$, we numerically test the maximal coherence measure $C_{m,n}$ that Gaussian evolution of either $|\psi_{N,m}\rangle$ or $|\phi_{N,n}\rangle$ produces. Explicitly, we determine the maxima
\begin{equation}
	\begin{aligned}
		T_{m,n}^{(N,n)}=\max_{|\phi_N\rangle,\xi,\alpha}C_{m,n}(|\xi,\alpha,\psi_{N,m}\rangle)\\
		T_{m,n}^{(N,m)}=\max_{|\psi_N\rangle,\xi,\alpha}C_{m,n}(|\xi,\alpha,\psi_{N,n}\rangle)
	\end{aligned}
\end{equation}
where $|\xi,\alpha,\psi_{N,m}\rangle=S(\xi)D(\alpha)|\psi_{N,m}\rangle$ and $|\xi,\alpha,\psi_{N,n}\rangle=S(\xi)D(\alpha)|\psi_{N,n}\rangle$. For given parameters $\xi$ and $\alpha$, we maximize over the states $|\psi_{N,m}\rangle$ or $|\psi_{N,n}\rangle$ employing the method in \cite{Fiurasek2022}, which allows us to obtain the maximizing states as an eigenvector, and therefore we can carry out the maximizing effeciently even for large $N$ since the numerical requirements are as hard as determining a root of a polynomial with degree $N$. Further, we determine the maximum over $\xi$ and $\alpha$ by random sampling and taking the maximal attempt. In these calculations, we inspect various examples to support a conjecture that $T_{m,n}^{(N,m)} \rightarrow 1$ and $T_{m,n}^{(N,n)} \rightarrow 1$ for large $N$.
To demonstrate the convergence, we focus on the cases with $m=0$ and $m=n-1$. Fig.~\ref{fig:SMConvergency} presents $1-T_{0,n}^{(N,l)}$ (with $l\in \{0,n\}$) and $1-T_{n-1,n}^{(N,l)}$ (with $l\in \{n-1,n\}$) for $n=4$ and $N \in \{5,...,10\}$. Also, the figure depicts both $1-T_{0,n}^{(N,l)}$ (with $l\in \{0,n\}$) and $1-T_{n-1,n}^{(N,l)}$ (with $l\in \{n-1,n\}$) for fixed $N$ and $n \in \{2,4,6,8\}$. It shows that the slowest convergence happens for the threshold $T_{0,8}^{(10,0)}$. While $1-T_{0,8}^{(N,0)}=0.264$, we obtain $1-T_{0,8}^{(20,0)}=2\times 10^{-4}$, which demonstrates clearly the tendency of $T_{0,8}^{(20,0)}$ to approach unity.

\section{Details on deriving a criterion employing the coherence measure $C_{0,1}$ together with the probabilities $P_1$ and $P_{e,1}$} \label{AppendixB}

Here, we provide technical details concerning the derivation of the criterion using the coherence measure $C_{0,1}$, the probability $P_1$ of one boson and the probability $P_{e,1}$ of more than one boson. First, we introduce the linear combination
\begin{equation}
	F_{\lambda_1,\lambda_2}(\rho)=C_{0,1}(\rho)+\lambda_1 P_1(\rho)+\lambda_2 P_{e,1}(\rho)
\end{equation}
with $\lambda_1$ and $\lambda_2$ free parameters. Then, we find the auxiliary maximum
\begin{equation}
	\widetilde{F}(\lambda_1,\lambda_2,k)\equiv \max_{\xi,\alpha}F_{\lambda_1,\lambda_2}(|\xi,\alpha,k\rangle),
\end{equation}
where $|\xi,\alpha,k\rangle=S(\xi)D(\alpha)|k\rangle$ corresponds to the Fock state $|k\rangle$ that undergoes the dynamics expressed as the action of the squeezing operator $S(\xi)$ and displacement operator $D(\alpha)$. Further, we derive the auxiliary function
\begin{equation}
	\widetilde{H}(\lambda_1,P_{e,1},k)\equiv \min_{\lambda_2}\left[ \widetilde{F}(\lambda_1,\lambda_2,k)-\lambda_2 P_{e,1}\right].
\end{equation}
A criterion follows as:
\begin{equation}\label{SM:minH}
	C_{0,1}(P_1,P_{e,1})>\min_{\lambda_1}\left[H(\lambda_1,P_{e,1})-\lambda_1 P_1\right],
\end{equation}
where $H(\lambda_1,P_{e,1})\equiv \max_k \widetilde{H}(\lambda_1,P_{e,1},k)$. The parameter $\lambda_1$ that minimizes (\ref{SM:minH}) requires one of the following:
\begin{equation}\label{SM:minEqs}
	\begin{aligned}
		\partial_{\lambda_1}H(\lambda_1,P_{e,1})&=P_1\\
		\widetilde{H}(\lambda_1,P_{e,1},0)&=\widetilde{H}(\lambda_1,P_{e,1},1).
	\end{aligned}
\end{equation}
If the first equation in Eq.~(\ref{SM:minEqs}) holds, the minimum in Eq.~(\ref{SM:minH}) corresponds to a stationary point. In case that the second equation in Eq.~(\ref{SM:minEqs}) is obeyed, the minimum in Eq.~(\ref{SM:minH}) happens for $\lambda_1$ where $H(\lambda_1,P_{e,1})$ is not smooth. Solving the minimum in Eq.~(\ref{SM:minH}) allows us to identify the criterion based on the knowledge of $C_{0,1}$, $P_1$ and $P_{e,1}$.

\end{appendix}

%\bibliography{hierarchiesCoherencesBib.bib}
\end{document}